\documentclass[aps,prl,twocolumn,superscriptaddress,showpacs,preprintnumbers]{revtex4}

\usepackage{graphicx}

\begin{document}

\title{Multiple Bosonic Mode Coupling in Electron Self-Energy of (La$_{2-x}$Sr$_x$)CuO$_4$}
\author{X. J. Zhou$^{1,2}$, Junren Shi$^{3}$, T. Yoshida$^{1,4}$, T. Cuk$^{1}$, W. L. Yang$^{1,2}$,
V. Brouet$^{1,2}$, J. Nakamura$^{1}$, N. Mannella$^{1,2}$, Seiki
Komiya$^{5}$, Yoichi Ando$^{5}$, F. Zhou$^{6}$, W. X. Ti$^{6}$, J.
W. Xiong$^{6}$,  Z. X. Zhao$^{6}$,  T. Sasagawa$^{1,7}$, T.
Kakeshita$^{8}$, H. Eisaki$^{1,8}$, S. Uchida$^{8}$, A.
Fujimori$^{4}$, Zhenyu Zhang$^{3,9}$, E. W. Plummer$^{3,9}$, R. B.
Laughlin$^{1}$,  Z. Hussain$^{2}$, and Z.-X. Shen
}

\affiliation{Dept. of Physics, Applied Physics and Stanford
Synchrotron Radiation Laboratory, Stanford University,  Stanford,
CA 94305
\\$^{2}$Advanced Light Source, Lawrence Berkeley National
Lab, Berkeley, CA 94720
\\$^{3}$Condensed Matter Sciences Division, Oak Ridge National Laboratory,
Oak Ridge, TN 37831
\\$^{4}$Dept. of Complexity Science and Engineering, University of
Tokyo, Kashiwa, Chiba 277-856, Japan
\\$^{5}$Central Research Institute of Electric Power Industry,
Komae, Tokyo 201-8511, Japan
\\$^{6}$National Lab for Superconductivity, Institute of Physics,
Chinese Academy of Sciences, Beijing 100080, China
\\$^{7}$Department of Advanced Materials Science, University of
Tokyo, Japan
\\$^{8}$Dept. of Superconductivity, University of Tokyo,
Bunkyo$-$ku, Tokyo 113, Japan
\\$^{9}$Department of Physics and Astronomy, University of Tennessee,
Knoxville, TN 37996
 }

\date{\today}

\begin{abstract}

High resolution angle-resolved photoemission spectroscopy data along
the (0,0)-($\pi$,$\pi$) nodal direction with significantly improved
statistics reveal fine structure in the electron self-energy of the
underdoped (La$_{2-x}$Sr$_x$)CuO$_4$ samples in the normal state.
Fine structure at energies of (40$\sim$46) meV and (58$\sim$63)meV,
and possible fine structure at energies of (23$\sim$29)meV and
(75$\sim$85)meV, have been identified. These observations indicate
that, in LSCO,  more than one bosonic modes are involved in the
coupling with electrons.

\end{abstract}


\maketitle

The recent observation of the electron self-energy renormalization
effect in the form of a  ``kink" in the dispersion has generated
considerable interest because it reveals a coupling of the electrons
with a collective boson mode of the cuprate
superconductors\cite{Damascelli}.  However, the nature of the bosons
involved remains controversial mainly because the previous
experiments can only be used to determine an approximate energy of
the mode and this energy is close to both the optical
phonon\cite{LanzaraEnergy,ZhouEnergy} and the spin
resonance\cite{KaminskiEnergy}. Determining the nature of the
mode(s) that couple to the electrons is likely important in
understanding the pairing mechanism of superconductivity.

In conventional superconductors, identification of the fine
structure for the phonon anomalies in the tunnelling spectra has
played a decisive role in reaching a consensus on the nature of the
bosons involved\cite{Rowell}. The fine structure provides
fingerprints for much more stringent comparison with known boson
spectra. So far, such fine structure has not been detected in the
angle-resolved photoemission spectroscopy (ARPES) data.  In this
Letter we present significantly improved high resolution ARPES data
of (La$_{2-x}$Sr$_x$)CuO$_4$ (LSCO) that, for the first time, reveal
fine structure in the electron self-energy, demonstrating the
involvement of multiple boson modes in the coupling with electrons.


\begin{figure}[tbp]
\begin{center}
\includegraphics[width=0.9\columnwidth,angle=0]{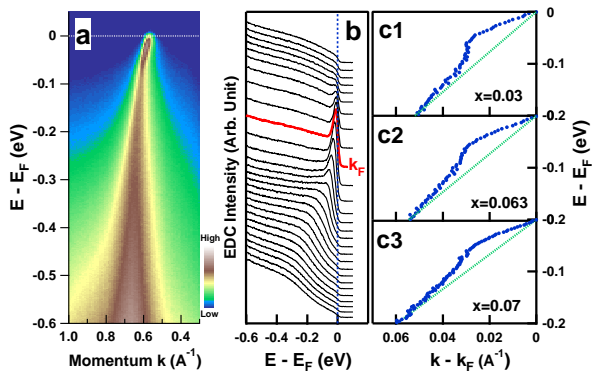}
\end{center}
\caption{(a) Raw data of a two-dimensional image showing the
photoelectron intensity as a function of momentum and energy for
LSCO x=0.063 sample. The intensity is represented by the
false-color. The measurement was taken along the (0,0)-($\pi$,$\pi$)
nodal direction at a temperature of $\sim$20 K. (b) The
photoemission spectra (energy distribution curves, EDCs) for the
LSCO x=0.063 sample corresponding to Fig. 1a. The spectrum at the
Fermi momentum k$_F$=(0.44$\pi/a$, 0.44$\pi/a$) (red curve) shows a
sharp peak. (c) The energy-momentum relation determined from MDC
method for LSCO x=0.03(c1), x=0.063 (c2) and x=0.07 (c3) samples.
The green dashed lines connecting the two points at the Fermi energy
and -0.2 eV are examples of a simple selection of the bare band.
 }
\end{figure}

The photoemission measurements were carried out on beamline 10.0.1
at the ALS, using Scienta 2002 and R4000 electron energy analyzers.
As high energy resolution and high data statistics are crucial to
identify fine structure in the electron self-energy, the
experimental conditions were set to compromise between these two
conflicting requirements. The measurement is particularly
challenging for LSCO system because of the necessity to use a
relatively high photon energy (55eV). Different energy resolution
between 12 and 20 meV was used for various measurements on different
samples, and the angular resolution is 0.3 degree. An example of the
high quality of the raw data is shown in Figs. 1a and 1b. Due to
space charge problem, the Fermi level calibration has a $\pm$5 meV
uncertainty. We mainly present our data on the heavily underdoped
LSCO x=0.03 (non-superconducting), LSCO x=0.063 (T$_c$=12 K) and
LSCO x=0.07 (T$_c$=14 K) samples. These heavily underdoped LSCO
samples are best candidates because they exhibit a stronger band
renormalization effect above T$_c$\cite{ZhouEnergy}; a relatively
large magnitude of the real self-energy makes the identification of
the fine structure easier. The LSCO single crystals are grown by the
travelling solvent floating zone method\cite{AndoSample}. The
samples were cleaved {\it in situ} in vacuum with a base pressure
better than 4$\times$10$^{-11}$ Torr. The measurement temperature
was $\sim$20K so all samples were measured in the normal state.

Fig. 1c shows the energy-momentum dispersion relation along the
(0,0)-($\pi$,$\pi$) nodal direction extracted by the MDC (momentum
distribution curves) method. Because of the larger band-width along
the nodal direction, the MDC method can be reliably used to extract
high quality data of dispersion in searching for fine structures. It
has also been shown theoretically that this approach is reasonable
in spite of the momentum-dependent coupling if we are only
interested in identifying the mode energies\cite{Sandvik}. As seen
in Fig. 1c, there is an abrupt slope change (``kink") in the
dispersions for LSCO samples with different dopings, similar to that
reported before\cite{ZhouEnergy}. However, the new data with
improved statistics indicates that the ``kink" has fine structure
and subtle curvatures in it, as seen for example in the LSCO x=0.03
sample (Fig. 1c1).

Since the bare dispersion is expected to be smooth in such a small
energy window, the ``kink" and its fine structure represent effects
associated with the electron self-energy. The real part of the
electron self-energy, Re$\Sigma$, can be extracted from the measured
dispersion (as in Fig. 1c) by subtracting the ``bare dispersion".
Following the convention\cite{Lashell,JRShi}, within a small energy
range near the Fermi level, the ``bare dispersion" can be assumed as
$\epsilon$$_0$(k)=$\alpha$$_1$(k-k$_F$)+$\alpha$$_2$(k-k$_F$)$^2$.
As we will describe later, the values of $\alpha$$_1$ and
$\alpha$$_2$ are determined so as to yield the best fit of the
measured dispersion from the maximum entropy method (MEM); the
choice of smooth bare band has little effect on the fine structure
that originate from the abrupt change in the dispersion. The
``effective" Re$\Sigma$ for LSCO at various doping levels is shown
in Fig. 2a. Also note the data were taken at different energy
resolutions, using different analyzers and under different
measurement conditions, as described in the caption of Fig. 2.

As seen from Fig. 2a, even with the most optimal experimental
conditions we can achieve and in these samples that exhibit the
strongest self-energy effects, there remains considerable noise in
the data statistics. However, by searching for peaks or curvature
changes, we can clearly identify some fine structure in the data.
Two clear features are around 40$\sim$46 meV and 58$\sim$63 meV, as
indicated by arrows in Fig. 2a, which show up clearly in x=0.03
(Fig. 2a1) and $\sim$0.06 samples (Fig. 2a4), although less clearly
in x=0.063 and x=0.07 samples. There may be another structure near
23$\sim$29 meV, which shows up mainly as a shoulder that is visible
in x=0.03 (Fig. 2a1), 0.063 (Fig. 2a2), 0.07 (Fig. 2a3) and possibly
in x=$\sim$0.06 sample (Fig. 2a4).  While these fine features are
subtle and one may argue about individual curves, the fact that we
have invariably observed them near similar energies in many
different samples and under different measurement conditions make
the presence of the fine structure convincing.

A natural question is whether the fine structure can be due to
instrumental artifact which may be related to detector inhomogeneity
and/or system noise. The detector problem can be ruled out because:
(1). The data were taken using the ``swept mode" of the electron
analyzer, i.e., each energy point was averaged over the entire
detector range along the energy direction. Therefore, all the energy
points for the same angle were taken under the same condition. (2).
The inhomogeneity along the angle direction is also minimized by
normalizing the photoemission spectra with the spectral weight above
the Fermi level which comes from the high harmonics of synchrotron
light and is angle-insensitive. (3). Among different measurements,
the corresponding detector angle with respect to the dispersion
bands vary due to slight changes in measuring geometry. The fact
that we observed consistent results suggest that the results are
intrinsic; (4). We have carried out our measurements using different
analyzers, from SES2002 to the latest R4000, yielding qualitatively
similar results. One can also rule out the possibility of noise
because they are supposed to be random. But for many different
measurements on different samples and under different experimental
conditions, the multiple structures are similar in energy within the
uncertainty caused by data statistics\cite{MomentumResolutionNote}.

The direct inspection in the raw data (Fig. 2a) has clearly
established the existence of fine structure in the electron
self-energy. This is independent of any models that are used in the
data analysis, including the MEM method we use below.  In order to
better quantify the characteristic energies of these fine features,
we take an approach to fit a smooth curve through the measured
Re$\Sigma$ data and then perform second-order derivative to the
fitted curve. Given the statistics level in the data, a spline
through the data is difficult because it is somewhat subjective. The
MEM procedure, which has been exploited to extract the spectral
features of the electron-phonon coupling from ARPES data for the
two-dimensional surface state of Be\cite{JRShi}, is well suited for
this purpose. By incorporating {\it prior} knowledge, including
positiveness of the bosonic function, and zero value of the bosonic
function at zero frequency and above a maximum frequency of 100meV,
the MEM is robust against the random noise in the data, insensitive
to the fitting details, and is therefore more objective\cite{JRShi}.
Since it remains unclear whether the method developed for
conventional metal can be extended to cuprate superconductors, as
the first step, we use it only as a procedure for curve fitting.

The fitted curves are shown in Fig. 2a together with the measured
Re$\Sigma$; the corresponding second-order derivative of the fitted
curves are shown in Fig 2b.  As expected, two dominant features near
40$\sim$46 meV and 58$\sim$63 meV, and one possible feature near
23$\sim$29, have been resolved in Fig. 2b. This is consistent with
the fine structure identified to a naked eye in the raw data of the
electron self-energy (Fig. 2a).  In addition, this data analysis
process also allows us to identify one more possible feature that
may exist at high energy near 75$\sim$85 meV, which seems to get
stronger with increasing doping. We note that, while the agreement
is not perfect at this stage, there is sufficient similarity to
suggest the detection of multiple modes.

\begin{figure}[tbp]
\begin{center}
\includegraphics[width=0.85\columnwidth,angle=0]{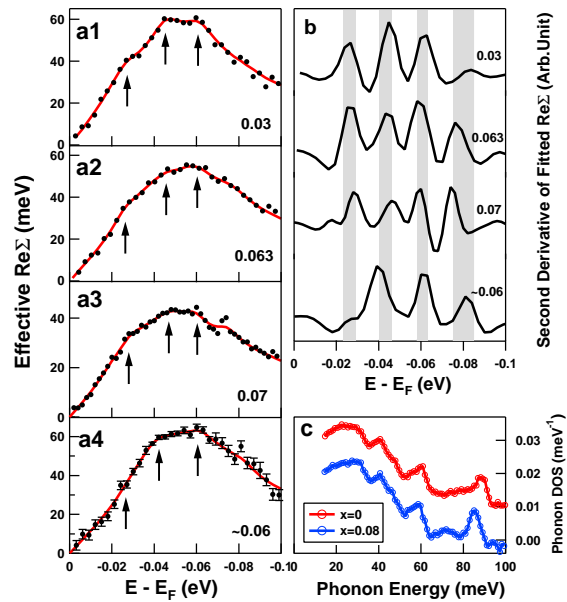}
\end{center}
\caption{(a).The effective real part of the electron self-energy for
LSCO x=0.03 (a1), 0.063 (a2), 0.07 (a3)and $\sim$0.06 (a4) samples.
Data (a1-a3) were taken using Scienta 2002 analyzer,  10eV pass
energy at an overall energy resolution (convoluted beamline and
analyzer resolution) of $\sim$18meV. Data (a4) were taken using
Scienta R4000 analyzer, 5eV pass energy at an overall energy
resolution of $\sim$12meV. For clarity, the error bar is only shown
for data (a4) which becomes larger with increasing binding energy.
The arrows in the figure mark possible fine structures in the
self-energy. The data are fitted using the maximum entropy method
(solid red lines). The values of ($\alpha$$_1$, $\alpha$$_2$) (the
unit of $\alpha$$_1$ and $\alpha$$_2$ are eV$\cdot$$\AA$ and
eV$\cdot$$\AA$$^2$, respectively) for bare band are (-4.25,0) for
(a1), (-4.25, 13) for (a2), (-3.7,7) for (a3) and (-4.3, 0) for
(a4). (b). The second-order derivative of the calculated Re$\Sigma$.
The ruggedness in the curves is due to limited discrete data points.
The four shaded areas correspond to energies of (23$\sim$29),
(40$\sim$46), (58$\sim$63) and (75$\sim$85) meV where the fine
features fall in. (c) The phonon density of state F($\omega$) for
LSCO x=0 (red) and x=0.08 (blue) measured from neutron
scattering\cite{McQueeney}. }
\end{figure}

The fine structure in the electron self-energy originates from the
underlying bosonic spectral function. The multiple features in Fig.
2b show marked difference from the magnetic excitation spectra
measured in LSCO which is mostly featureless and doping
dependent\cite{HaydenTranquada}. In comparison, the features in Fig.
2b show more resemblance to the phonon density-of-states (DOS),
measured from neutron scattering on LSCO (Fig. 2c)\cite{McQueeney},
in the sense of the number of modes and their positions. This
similarity between the extracted fine structure and the measured
phonon features favors phonons as the nature of bosons involved in
the coupling with electrons. In this case, in addition to the
half-breathing mode at 70$\sim$80 meV that we previously considered
strongly coupled to electrons\cite{LanzaraEnergy}, the present
results suggest that several lower energy optical phonons of oxygens
are also actively involved.

\begin{figure}[tbp]
\begin{center}
\includegraphics[width=0.9\columnwidth,angle=0]{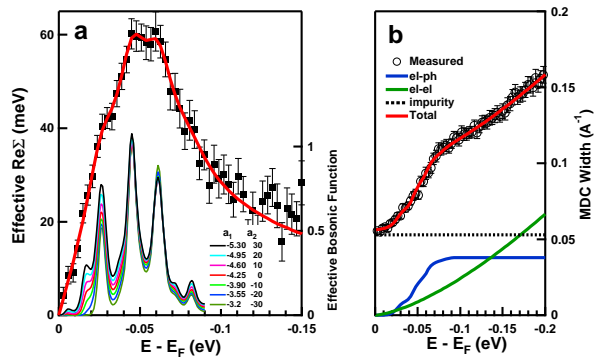}
\end{center}
\caption{(a)Real part of the electron self-energy for LSCO x=0.03
 as obtained from the dispersion shown in Fig. 1c1 (solid square) and
calculated from the extracted effective bosonic spectral using the
MEM procedure (red solid curve) with $\alpha$$_1$=-4.25 and
$\alpha$$_2$=0. Also plotted are the effective bosonic functions
obtained by using different bare bands as represented by the
different sets of $\alpha$$_1$ and $\alpha$$_2$ values. (b) The MDC
width of LSCO x=0.03 (open circles). The contribution from the
electron-phonon coupling (blue line) is calculated from the
effective bosonic function in Fig. 3a with $\alpha$$_1$=-4.25 and
$\alpha$$_2$=0. The ``impurity" contribution is assumed to be a
constant, 0.053 $\AA$$^{-1}$ (dotted black line). The momentum
resolution here is 0.019 $\AA$$^{-1}$. After subtracting all the
electron-phonon and ``impurity" contributions, the residual part is
fitted by C$\omega$$^{\alpha}$ with C$\sim$0.7 and $\alpha$$\sim$1.5
(green line). }
\end{figure}

For conventional metals, the MEM procedure can be used to extract
the Eliashberg function which gives spectral features of the
electron-phonon coupling\cite{JRShi}. For strongly correlated
cuprate superconductors, {\it a priori} it is unclear whether the
Eliashberg formalism is applicable or not. However, we note that the
nodal excitation of LSCO in the normal state may provide a closest
case for the procedure to be applicable. Recent transport
measurements reveal that a quasiparticle picture may be still
reasonable for electrons near the nodal direction, even for very low
doping\cite{AndoTransport}. This is consistent with the ARPES data
which show a well-defined peak in the nodal spectra in the lightly
doped samples\cite{ZhouEnergy}. The nodal dispersion does not have
pronounced curvature in its bare band, unlike the strong curvature
near the saddle point of the antinodal region. The selection of the
nodal direction in the normal state also minimizes complications due
to the existence of a superconducting gap or pseudogap in the
extraction process.

Given these considerations and the fact that there is no better
alternative available, we made an attempt by applying the Eliashberg
formalism and the MEM procedure to extract the effective bosonic
function from the real part of the electron self-energy (Fig. 3a).
It is clear that the multiple features are rather robust against the
choice of the bare band by varying $\alpha$$_1$ and $\alpha$$_2$.
All other tests as detailed in \cite{JRShi} have been carried out.
The fine structure as obtained from LSCO x=0.03 is in agreement with
that in the second-order derivative shown in Fig. 2b. The calculated
real part of the electron self-energy is plotted in Fig. 3a together
with the measured data. Fig. 3b shows the MDC width which is
directly related to Im$\Sigma$=($\Gamma$/2)$v_0$, with $\Gamma$
being the MDC width (Full-Width-at-Half-Maximum, FWHM) and $v_0$ the
bare velocity. While there is a drop near 75 meV, there is an
overall increase of the MDC width with increasing binding energy
(Fig. 3b), which is different from simple electron-phonon coupling
systems such as Be\cite{Lashell}. The MEM analysis allows us to
calculate the contribution of the electron-phonon coupling which
gives rise to the abrupt drop in Im$\Sigma$. We note that this
calculation has some uncertainty related to the bare band selection
(Fig. 3a).  After subtracting the contributions from the
electron-phonon coupling, ``impurity" scattering,  and angular
resolution, the residual part is found to be proportional to
$\omega$$^{\alpha}$(Fig. 3b).  This term most likely represents the
contribution of the electron-electron interaction. The corresponding
electron-electron contribution in the real part of the self-energy
is a smooth function and may be absorbed into the ``bare dispersion"
because we focus only on abrupt structure in Re$\Sigma$ in
extracting the bosonic spectral function. Here we also note that
while the imaginary part of the electron self-energy is consistent
with the existence of electron-phonon coupling, it is difficult to
identify the fine structure as has been done for the real part
because of the larger experimental uncertainty in determining the
peak width over the peak position. This analysis also shows that
there is an internal consistency in the MEM procedure that connects
the real and imaginary parts of the self-energy.

In summary, by taking high resolution data on heavily underdoped
LSCO samples with high statistics, we have detected fine structure
in the electron self-energy.  This indicates multiple bosonic modes
are involved in the coupling to electrons in the LSCO system.

The work at the ALS and SSRL is supported by the DOE's Office of
BES, Division of Material Science, with contract
DE-FG03-01ER45929-A001 and DE-AC03-765F00515. The work at Stanford
was also supported by NSF grant DMR-0304981 and ONR grant
N00014-04-1-0048-P00002. EWP is supported by DOE DMS and
NSF-DMR-0451163. The work at Oak Ridge National Laboratory was
partially supported through DOE under Contract DE-AC05-00OR22725.
The work in Japan is supported by a Grant-in-Aid from the Ministry
of Education, Culture, Sports, Science and Technology of Japan and
the NEDO. The work in China is supported by NSF of China and
Ministry of Science and Technology of China through Project 10174090
and Project G1999064601.


\begin{thebibliography}{99}

\bibitem{Damascelli} A. Damascelli, Z.-X. Shen and Z. Hussain,
Rev. Mod. Phys. 75 (2003) 473.
\bibitem{LanzaraEnergy} A. Lanzara et al., Nature (London) {\bf
412}, 510(2001).
\bibitem{ZhouEnergy} X. J. Zhou et al., Nature (London) {\bf 423},
398 (2003); X. J. Zhou, et al., Phys. Rev. Lett. {\bf 92}
187001(2004); T. Yoshida et al., Phys. Rev. Lett. {\bf 91}, 027001
(2003).
\bibitem{KaminskiEnergy} A. Kaminski et al., Phys. Rev. Lett.  {\bf 86}, 1070
(2001); P. D. Johnson et al., Phys. Rev. Lett.  {\bf 87}, 177007
(2001); S. V. Borisenko et al., Phys. Rev. Lett. {\bf 90}, 207001
(2003).
\bibitem{Rowell} J. M. Rowell et al.,Phys. Rev. Lett. {\bf 10}, 334
(1963); D. J. Scalapino et al., Phys. Rev. {\bf 148}, 263 (1966).
\bibitem{AndoSample} S. Komiya et al., Phys. Rev. B {\bf 65}, 214535 (2002);
F. Zhou et al., Supercon. Sci. Technol. {\bf 16}, L7 (2003).
\bibitem{Sandvik} A. W. Sandvik et al., cond-mat/0309171; T. P.
Devereaux et al., Phys. Rev. Lett. {\bf 93},117004 (2004).
\bibitem{Lashell}S. Lashell et al., Phys.  Rev. B  {\bf 61}, 2371(2000).
\bibitem{MomentumResolutionNote}From the simulations we have done,
a finite momentum resolution has little effect on the fine
structures in the dispersion extracted from the MDC method. High
energy resolution is important to identify the fine structures.
\bibitem{JRShi} J. R. Shi et al., Phys. Rev. Lett. {\bf 92},
186401 (2004).
\bibitem{HaydenTranquada}The spin excitation spectrum of LSCO x=0.14
shows a broad peak at a lower energy ($\sim$20 meV) (S. M. Hayden
et al., Phys. Rev. Lett. {\bf 76}, 1344 (1996)).  This peak is
pushed down to below 5 meV in LSCO x=0.07 (H. Hiraka et al., J.
Phys. Soc. Jpan {\bf 70}, 853 (2001)) and in x=0.05 (H. Goka
Physica C {\bf 388-389}, 239(2003)).  A measurement from stripe
ordered La$_{1.875}$Ba$_{0.125}$CuO$_4$ and related calculation
showed a broad feature centered around 50$\sim$60 meV (Tranquada
et al., Nature {\bf 429}(2004)534).  Given that this is a
different material at different doping, we do not consider this as
relevant to the LSCO x=0.03$\sim$0.07. The rapid change of spin
spectra with doping has also been observed in
YBa$_2$Cu$_3$O$_{7-\delta}$, S. Chakravarty et al.,  Phys. Rev. B
{\bf 63}, 094503 (2001)).
\bibitem{McQueeney} R. J. McQueeney et al., Phys. Rev. Lett. {\bf 87}, 077001
(2001);L. Pintschovious and M. Braden, Phys. Rev. B {\bf 60}, R15039
(1999).
\bibitem{AndoTransport} Y. Ando et al., Phys. Rev. Lett. {\bf 87}, 017001 (2001).


\end{thebibliography}
\end{document}